\documentclass[aps, prl, twocolumn, superscriptaddress]{revtex4-2}

\usepackage[]{amsmath}
\usepackage{amssymb}
\usepackage{mathrsfs}

\usepackage{fontawesome}

\usepackage[]{graphicx}
\graphicspath{{figures/}}
\usepackage[]{mathtools}
\usepackage[]{bm}
\usepackage[]{braket}
\usepackage{esint}
\usepackage{ulem}
\usepackage{cancel}
\usepackage[caption=false]{subfig}

\newcommand{\figref}[1]{\figurename~\ref{#1}}

\newcommand{\gl}{{\raisebox{.15em}[0em][0em]{$\scriptscriptstyle{>}$}\hspace{-.52em}\raisebox{-.15em}[0em][0em]{$\scriptscriptstyle{<}$}}}
\newcommand{\ssl}{{\scriptscriptstyle{<}}}
\newcommand{\ssg}{{\scriptscriptstyle{>}}}
\newcommand{\ssm}{{\scriptscriptstyle{-}}}
\newcommand{\ssp}{{\scriptscriptstyle{+}}}
\newcommand{\Inc}{\ssm\ssg}
\newcommand{\Refl}{\ssp\ssg}
\newcommand{\Tra}{\ssm\ssl}

\usepackage{cases}
\usepackage[]{comment}

\begin{document}

\title{
  \v{C}erenkov radiation in vacuum from a superluminal grating
}

\author{Daigo Oue}
\affiliation{The Blackett Laboratory, Department of Physics, Imperial College London, Prince Consort Road, Kensington, London SW7 2AZ, United Kingdom}
\author{Kun Ding}
\affiliation{The Blackett Laboratory, Department of Physics, Imperial College London, Prince Consort Road, Kensington, London SW7 2AZ, United Kingdom}
\affiliation{Department of Physics, State Key Laboratory of Surface Physics, and Key Laboratory of Micro and Nano Photonic Structures (Ministry of Education), Fudan University, Shanghai 200438, China}
\author{J. B. Pendry}
\affiliation{The Blackett Laboratory, Department of Physics, Imperial College London, Prince Consort Road, Kensington, London SW7 2AZ, United Kingdom}

\date{\today}

\begin{abstract}
  Nothing can physically travel faster than light in vacuum.
  This is why it has been considered that there is no \v{C}erenkov radiation (\v{C}R) without an effective refractive index due to some background field.
  In this Letter,
  we theoretically predict \v{C}R in vacuum from a spatiotemporally modulated boundary.
  We consider the modulation of traveling wave type and apply a uniform electrostatic field on the boundary to generate electric dipoles.
  Since the induced dipoles stick to the interface,
  they travel at the modulation speed.
  When the grating travels faster than light,
  it emits \v{C}R.
  In order to quantitatively examine this argument,
  we need to calculate the field scattered at the boundary.
  We utilise a dynamical differential method,
  which we developed in the previous paper,
  to quantitatively evaluate the field distribution in such a situation.
  We can confirm that all scattered fields are evanescent if the modulation speed is slower than light while some become propagating if the modulation is faster than light.
\end{abstract}

\maketitle

\textit{Introduction.}---%
\v{C}erenkov radiation (\v{C}R) is radiation from a charged particle moving faster than light in a medium,
which was originally observed by \v{C}erenkov in 1934 \cite{cherenkov1934visible} and then theoretically studied by Tamm and Frank \cite{frank1937coherent}.
It has been observed in various systems,
including metamaterials and photonic crystals where the speed of light is effectively suppressed and hence the threshold for \v{C}R
\cite{%
  pendry1994energy,%
  antipov2008observation,%
  xi2009experimental,%
  tao2016reverse%
}.
There is also \v{C}R into surface modes such as surface plasmon polaritons and Dyankov waves 
\cite{%
  liu2012surface,%
  fares2019quantum,%
  hao2020surface%
}

Even uncharged moving particles can emit \v{C}R if electrically or magnetically polarised,
which is closely related to the friction induced by electromagnetic fields
\cite{%
  pendry1998can,%
  maghrebi2013quantum,%
  milton2020self%
}.
Many attempts to generate \v{C}R in linear optics were focused on slowing down the speed of light by engineering the medium's dispersion relation as in the case of photonic crystals and metamaterials.
On the other hand, 
it has been reported that \v{C}R is emitted not only from physically moving dipoles but also from ones induced, for example, by moving light foci 
\cite{%
  auston1984cherenkov,%
  bakunov2009cherenkov,%
  smith2016steerable%
}
and solitons in nonlinear optical fibres 
\cite{%
  cao1994soliton,%
  akhmediev1995cherenkov,%
  chang2010highly%
}
or micro resonators
\cite{%
  brasch2016photonic,%
  cherenkov2017dissipative,%
  vladimirov2018effect%
}.

\v{C}R is composed of coherent multi-frequency components propagating in the same direction.
It is for this property that \v{C}R has not only attracted scientific interest but also been applied to other research fields.
\v{C}R emitted by cosmic rays has been captured in large facilities such as super-Kamiokande and played vital roles in astrophysics and high energy physics \cite{fukuda2003super}.
As mentioned above,
\v{C}R can be generated in nonlinear media.
It is utilised as an optical frequency comb, 
as a tuneable and broadband source and is expected to be utilised in spectroscopy or metrology
\cite{%
  tu2009optical,%
  skryabin2010colloquium,%
  brasch2016photonic%
}.

There are many reports of \v{C}R in media as reviewed above;
however,
there is a limited number of studies on \v{C}R in vacuum.
This is because nothing can physically move faster than light in vacuum.
Previous studies show that introducing some background such as external electromagnetic or Chern-Simons fields induces effective refractive index to lower the \v{C}R threshold in vacuum below the light barrier
\cite{%
  kaufhold2007vacuum,%
  macleod2019cherenkov,%
  lee2020cherenkov,%
  artemenko2020quasiclassical%
}.

\begin{figure}[htbp]
  \centering
  \includegraphics[width=.9\linewidth]{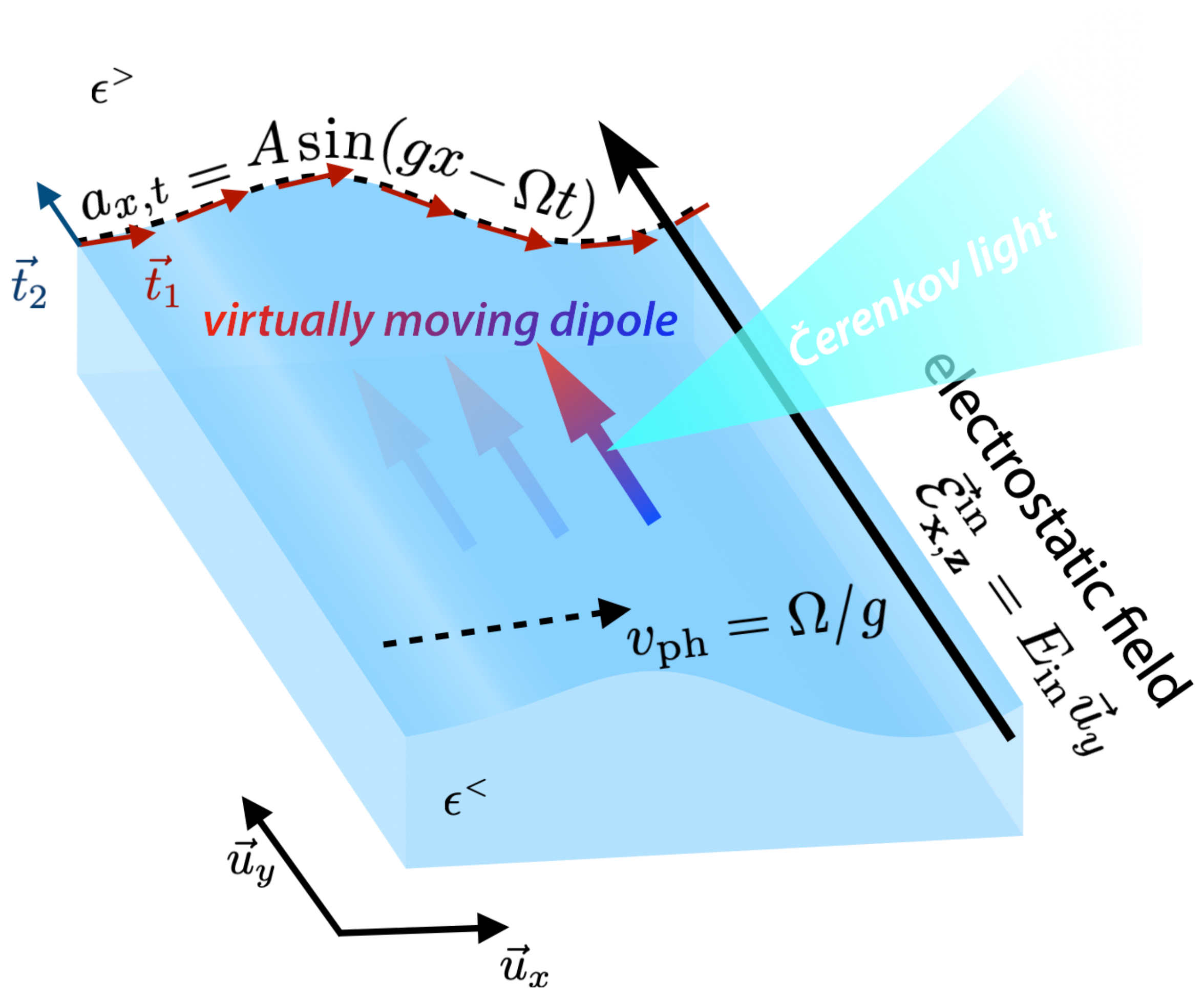}
  \caption{
    Single interface system composed of a dielectric and vacuum.
    The interface is under the spatiotemporal modulation of traveling wave type \eqref{eq:a}.
    The orthonormal tangential vectors, $\vec{t}_{1,2}$ can be calculated from the interface profile.
    The permittivities above and below the interface are denoted as $\epsilon^\ssg$ and $\epsilon^\ssl$,
    respectively.
    We apply a uniform electrostatic field 
    $\vec{\mathcal{E}}_{\mathbf{x},z}^\mathrm{in}$
    on the modulated interface so that there are induced dipoles which travel on the interface due to its profile of the traveling wave type.
    The velocity of the grating $v_\mathrm{ph} = \Omega/g$ can be tuned by two independent parameters so that it can exceed the speed of light.
    When the speed is faster than light,
    the induced dipoles emit \v{C}erenkov radiation.
  }
  \label{fig:fig1}
\end{figure}

In this Letter,
we propose a mechanism for \v{C}R in vacuum without introducing any effective index.
We consider a single interface system composed of vacuum and a dielectric (\figref{fig:fig1}).
By spatiotemporally modulating its interface,
we can generate an interface profile of a travelling wave type,
\begin{align}
  a_\mathbf{x} = A\sin (\mathbf{q}\cdot\mathbf{x}) = A\sin\left[g\left(x-\frac{\Omega}{g}t\right)\right],
  \label{eq:a}
\end{align}
where we have defined a three-component vector $\mathbf{x} \equiv \{x,y,ict\}$ and a reciprocal vector $\mathbf{q} \equiv \{g,0,i\Omega/c\}$ for the sake of convenience,
and $v_\mathrm{ph} \equiv \Omega/g$ is the sliding speed of the profile.
Note that $\Omega$ and $g$ are independent modulation parameters (temporal and spatial modulation frequencies),
and thus the sliding speed $v_\mathrm{ph}$ is not limited by the speed of light.
Since we are focusing on dielectrics,
the permeability is assumed to be unity everywhere ($\mu = 1$).
The permittivity is given by means of the interface profile,
\begin{align}
  \epsilon_{\mathbf{x},z} 
  = \alpha \Theta(a_\mathbf{x}-z) + \epsilon^\ssg,
\end{align}
where we denote the permittivities of the upper and lower media as $\epsilon^\gl$,
the permittivity difference as $\alpha \equiv \epsilon^\ssl - \epsilon^\ssg$
and the Heaviside unit step function as $\Theta(z)$.

When an electrostatic field 
$\vec{\mathcal{E}}^\mathrm{in} = E_\mathrm{in} \vec{u}_y$
is applied to this configuration,
electric dipoles are induced on the interface.
Following the traveling type profile of the interface,
the induced dipoles move virtually along the interface.
If the profile velocity and hence that of the induced dipoles is faster than light,
the induced dipoles may emit \v{C}R.

Our system is periodic in the $x$ direction and in time,
and thus the radiation wavenumber in the $x$ direction and the frequency in our system are written 
$k_{x,m} = k_x + mg$ and $\omega_m = \omega + m\Omega$
in accordance with the Floquet-Bloch theory,
where $m = 0, \pm 1, \pm 2, \cdots$ is the diffraction order.
Note that here we have multi-frequency radiation in contrast with ordinary diffraction.
The propagarion direction of the $m$th order diffraction is $\theta_m^\tau = \cos^{-1} (ck_{x,m}/\omega_m\sqrt{\epsilon^\tau})$ in each medium.
Since we have a uniform electrostatic field ($k_x = 0$, $\omega = 0$),
all the polychromatic waves in the different diffraction orders propagate in the same direction in each medium,
$\theta_m^\tau \rightarrow \cos^{-1} (c/v_\mathrm{ph}\sqrt{\epsilon^\tau}) \equiv \theta_{\scriptscriptstyle{\mathrm{\check{C}R}}}^\tau$.
We quantitatively analyse such a situation below.

\textit{Modulation-induced source at the interface.}---%
Here, we briefly review a dynamical differential method,
which we developed in the previous work \cite{oue2021calculating},
and calculate the diffraction at the interface subject to modulations in space and time.
Our previous work is based on the coordinate translation method originally proposed by Chandezon et al.
\cite{%
  chandezon1980new,%
  chandezon1982multicoated%
},
where they consider static corrugated interfaces and match Maxwell's boundary conditions directly at the interfaces with the help of differential geometry.
Using this method,
it is straightforward to take the structure of the interfaces into consideration,
and it has been utilised to calculate structured surfaces of various media,
including anisotropic, plasmonic and dielectric materials
\cite{%
  harris1995differential,%
  barnes1995photonic,%
  kitson1996surface,%
  kitamura2013hermitian,%
  murtaza2017study%
}
It is also worth noting that there is a series of studies,
which confirm that the method works well for smooth shallow corrugations and propose possible ways to improve the method so that they can handle deep corrugation even with sharp edges
\cite{%
  xu2014simple,%
  xu2017numerical,%
  xu2020numerical,%
  shcherbakov2013efficient,%
  shcherbakov2017generalized,%
  shcherbakov2019curvilinear,%
  essig2010generation%
}.
In these works,
local distortion of the coordinate systems is applied instead of the global translation of the coordinate in order to improve the convergence.
Since we are interested in the time-dependent corrugation of the traveling wave type,
we assume that the corrugation depth is small and safely work on the simple global coordinate translation.

\begin{figure}[htbp]
  \centering
  \includegraphics[width=.7\linewidth]{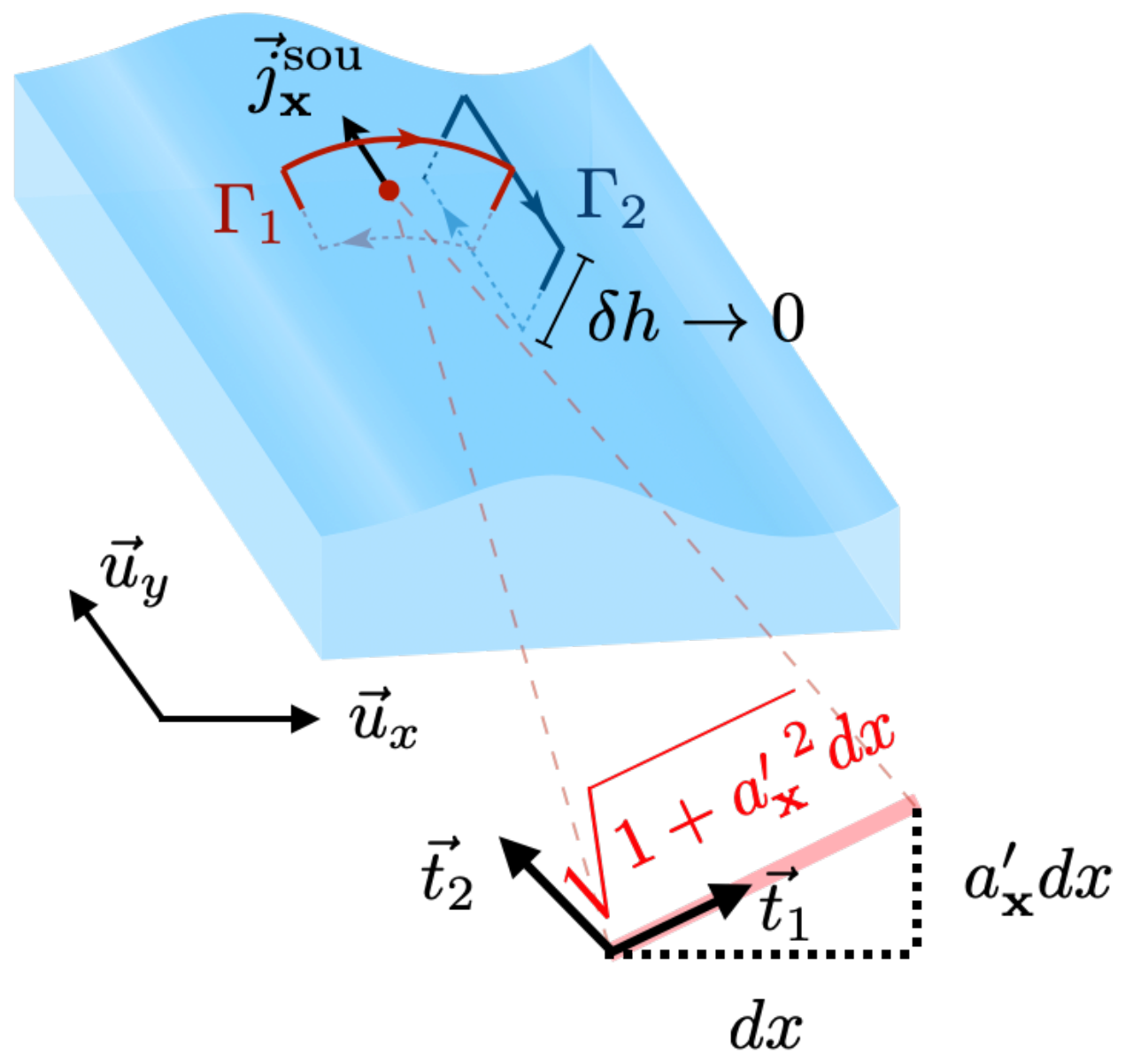}
  \caption{
    Integration paths leading to the boundary conditions at the modulated interface.
    The modulation induced source, 
    $\vec{j}_{\mathbf{x}}^\mathrm{sou}$,
    comes into contribution.
  }
  \label{fig:tddf}
\end{figure}

In order to calculate diffraction at the structured interface,
we need relevant boundary conditions.
By integrating Maxwell-Heaviside equations over two kinds of path,
$\Gamma_{1,2}$,
which enclose the interface as shown in \figref{fig:tddf},
we can derive the boundary conditions,
\begin{align}
  \begin{cases}{}
  \eta\vec{t}_1 
  \cdot
  \left(
    \vec{\mathcal{H}}_{\mathbf{x},a_\mathbf{x}+0}
    -\vec{\mathcal{H}}_{\mathbf{x},a_\mathbf{x}-0}
  \right)
  &=
  \vec{t}_2 \cdot \vec{j}_{\mathbf{x}}^\mathrm{sou}
  \\
  \vec{t}_2 
  \cdot
  \left(
    \vec{\mathcal{E}}_{\mathbf{x},a_\mathbf{x}+0}
    -\vec{\mathcal{E}}_{\mathbf{x},a_\mathbf{x}-0}
  \right)
  &= 0.
  \end{cases}
  \label{eq:bc}
\end{align}
Here, 
we have a modulation-induced source term,
\begin{align}
  \vec{j}_{\mathbf{x}}^\mathrm{sou}
  &=
  \cfrac{\dot{a}_\mathbf{x}}{c}\alpha 
  \vec{\mathcal{E}}_{\mathbf{x},a_\mathbf{x}-0},
\end{align}
which is finite if there is time dependence (i.e. $\dot{a}_\mathbf{x} \neq 0$).
This source term is responsible for radiation from the interface.
Note that we take an unit system such that $Z_0 \equiv \sqrt{\epsilon_0}{\mu_0} = 1$ for simplicity.
The orthonormal tangential vectors of the interface, $\vec{t}_{1,2}$, can be given by means of the interface profile,
\begin{align}
      \vec{t}_{1}
    =
    \cfrac{\vec{u}_x + a_\mathbf{x}' \vec{u}_z}{\sqrt{1+{a_\mathbf{x}'}^2}},
    \quad
    \vec{t}_2
    =
    \vec{u}_y,
\end{align}
and the electric and magnetic fields in real space as $\vec{\mathcal{E}}_{\mathbf{x},z}$ and $\vec{\mathcal{H}}_{\mathbf{x},z}$. 
Note that we have a factor of $\eta = \sqrt{1+{a_\mathbf{x}'}^2}$ in front of $\vec{t}_1$ in Eq. \eqref{eq:bc},
which corresponds to the spatial variation (expansion or contraction) of the differential line element due to the coordinate translation,
while we do not have that factor at $\vec{t}_2$ because the translation does not affect the line element in the direction.

Since our system is periodic in space and time,
we can substitute Floquet-Bloch type solutions in the upper and lower media,
e.g. $
\vec{\mathcal{E}}_{\mathbf{x},z} = 
\sum_m E_{\mathbf{k}_m,z} \vec{t}_2 e^{i\mathbf{k}_n\cdot\mathbf{x}}\
(\mathbf{k}_m = \mathbf{k} + m\mathbf{q},\ \mathbf{k} \equiv \{k_x, 0, ik_0 = i\omega/c\})
$,
to obtain simultaneous equations in the reciprocal space,
\begin{align}
  \begin{pmatrix}
    \mathsf{N}_{\mathbf{k}}^{\ssp\ssg} & -(\mathsf{N}_{\mathbf{k}}^{\ssm\ssl} + \mathsf{L}_{\mathbf{k}})\\
    \mathsf{M}_{\mathbf{k}}^{\ssp\ssg} & -\mathsf{M}_{\mathbf{k}}^{\ssm\ssl}
  \end{pmatrix}
  \begin{pmatrix}
    \mathbb{E}_{\mathbf{k}}^{\ssm\ssl}\\
    \mathbb{E}_{\mathbf{k}}^{\ssp\ssg}
  \end{pmatrix}
  =
  \begin{pmatrix}
     -\mathsf{N}_{\mathbf{k}}^{\ssm\ssg}
    \mathbb{E}_{\mathbf{k}}^{\ssm\ssg}\\
     -\mathsf{M}_{\mathbf{k}}^{\ssm\ssg}
    \mathbb{E}_{\mathbf{k}}^{\ssm\ssg}
  \end{pmatrix}
\end{align}
where we collect the modal amplitude in each diffraction order to form
${\mathbb{E}_{\mathbf{k}}^{\sigma\tau}} 
= 
\begin{pmatrix}
  \cdots
  &
  E_{\mathbf{k}_{-1}}^{\sigma\tau}
  &
  E_{\mathbf{k}_{0}}^{\sigma\tau}
  &
  E_{\mathbf{k}_{+1}}^{\sigma\tau}
  &
  \cdots
\end{pmatrix}^\intercal$.
We have specified the upper and lower media by $\tau = \gl$,
and the propagating direction is labeled by $\sigma = \pm$.
Thus, we can regard $E_{\mathbf{k}_m}^{\Inc}$ as the incident components from the upper side while $E_{\mathbf{k}_m}^{\Refl}$ and $E_{\mathbf{k}_m}^{\Tra}$ as the diffracted components in the upper and lower sides.
The effects of the source term and interface geometry are encoded in the $\mathsf{L}$, $\mathsf{M}$ and $\mathsf{N}$ coefficients matrices,
whose elements read
\begin{align}
  [\mathsf{L}_{\mathbf{k}}]_{lm}
  &=
  \frac{\alpha}{c}
  \frac{(l-m)\Omega}{-K_{\mathbf{k}_m}^{\ssl}}
  \times
  \operatorname{sgn}(\omega_m)
  \frac{k_{x,m}}{|k_{x,m}|}
  J_{l-m}(\phi_{\mathbf{k}_m}^{\ssm\ssl}),
  \label{eq:L}
  \\
  [\mathsf{M}_{\mathbf{k}}^{\sigma\tau}]_{lm}
  &=
  \frac{k_{x,m}}{|k_{x,m}|}
  \operatorname{sgn}(\omega_m)
  J_{l-m}(\phi_{\mathbf{k}_m}^{\sigma\tau}),
  \label{eq:M}
  \\
  [\mathsf{N}_{\mathbf{k}}^{\sigma\tau}]_{lm}
  &=
  \left(
    \frac{\sigma K_{\mathbf{k}_m}^\tau}{k_{x,m}}
    -
    \frac{(l-m)g}{\sigma K_{\mathbf{k}_m}^\tau}
  \right)
  \frac{|k_{x,m}|}{|k_{0,m}|}
  J_{l-m}(\phi_{\mathbf{k}_m}^{\sigma\tau}).
  \label{eq:N}
\end{align}
Here, 
$J_m$ is the $m$th order Bessel function of the first kind,
and we have defined the wavenumber in the $z$ direction,
\begin{align}
K_{\mathbf{k}} 
\equiv 
\operatorname{sgn}(\omega)\operatorname{Re}\sqrt{\epsilon^\tau {k_0}^2 - {k_x}^2}
+
i\operatorname{Im}\sqrt{\epsilon^\tau{k_0}^2 - {k_x}^2},
\end{align}
and the corresponding propagating phase $\phi_{\mathbf{k}_m}^{\sigma \tau} \equiv \sigma K_{\mathbf{k}_m}^\tau A$.
By truncating $\mathsf{M}, \mathsf{N}$ and $\mathsf{L}$ matrices to finite rank ones (i.e. $|l|,|m| \leq m_c$),
we can numerically invert the matrix on the left hand side and evaluate the diffraction amplitudes.

\begin{figure*}[htbp]
  \centering
  \includegraphics[width=\linewidth]{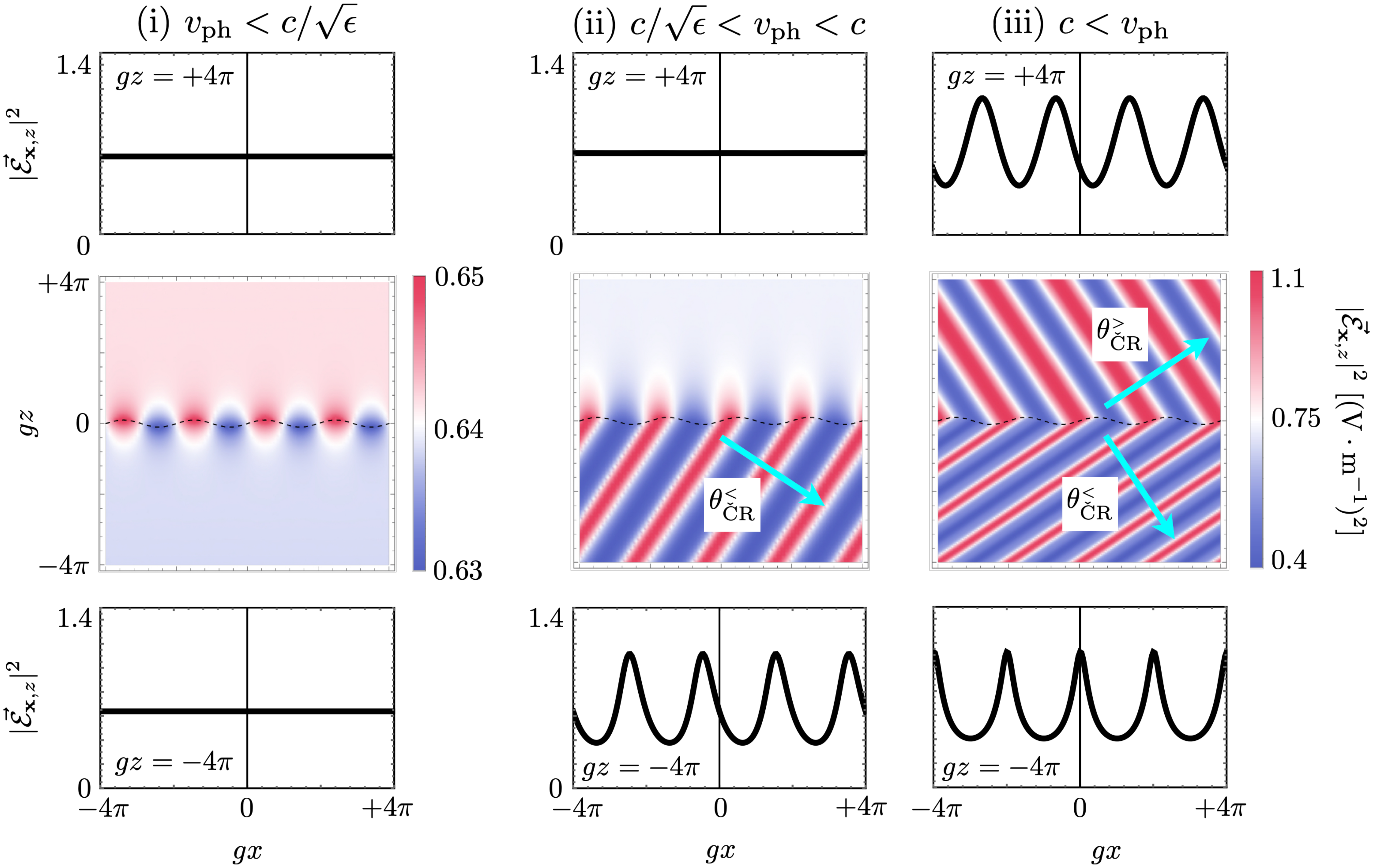}
  \caption{
    Snapshots (middle row) and their cross sections (upper and lower rows) of the field distributions in the subluminal and superluminal regimes.
    The intensity of the total electric field $|\vec{\mathcal{E}}_{\mathbf{x},z}|^2$ is plotted.
    (i) 
    Subluminal regime in the dielectric and vacuum sides,
    $v_\mathrm{ph} = \Omega/g = 0.2c < c/\sqrt{\epsilon}$.
    (ii)
    Superluminal in the dielectric side and subluminal in the vacuum side,
    $c/\sqrt{\epsilon} < v_\mathrm{ph} = 0.8c < c$.
    (iii) 
    Superluminal regime in both sides,
    $c < v_\mathrm{ph} = 1.2c$.
    The modulation parameters are given as following:
    the spatial frequency $g = 2\pi\ [\mathrm{\mu m^{-1}}]$,
    the modulation depth $2A = 100\ [\mathrm{nm}]$.
    The input amplitude is $E_\mathrm{in} = 1\ \mathrm{[V\cdot m^{-1}]}$.
    The permittivities are $\epsilon^\ssg = 1$ and $\epsilon^\ssl = \epsilon = 2.25$.
    The cutoff is $m_c = 10$ so that we take $2m_c+1 = 21$ diffracted waves into account. 
    The colorbars for (ii) and (iii) are common and shown on the right of (iii).
    Note that the scale of the colorbar of (i) is different from those of (ii) and (iii),
    and the horizontal axes and the vertical axes of the color plots are normalised by the spatial period $g$ of the modulation.
  }
  \label{fig:field}
\end{figure*}

\textit{\v{C}erenkov radiation.}---%
We reconstruct the diffracted field distribution in real space after obtaining the diffraction amplitudes.
Our interest is to apply a uniform electrostatic field ($\omega = 0, k_x = 0$).
In \figref{fig:field}, 
we show snapshots of the field distributions for various modulation speeds and the corresponding cross section plots in the far field region. 
When the modulation speed is slower than that of light in the dielectric (subluminal regime),
the distribution is uniform in the far field region both on the dielectric and vacuum sides.
This is because,
in the subluminal regime ($v_\mathrm{ph} = \Omega/g < c/\sqrt{\epsilon}$),
the wavenumber in the $z$ direction is imaginary for each diffraction order,
\begin{align}
  K_{\mathbf{k}_m}^\tau 
  = \frac{|m|g\sqrt{\epsilon^\tau}}{c}
  i \operatorname{Im}
  \sqrt{{v_\mathrm{ph}}^2 - \left(\frac{c}{\sqrt{\epsilon^\tau}}\right)^2},
\end{align}
and hence all diffracted fields exponentially decay.

On the other hand,
once the modulation speed is faster than light in the dielectric and vacuum (superluminal regime),
we can find the \v{C}erenkov type patterns (i.e. propagating waves) in each medium.
This is because,
unlike in the subluminal regime,
the wavenumber in the $z$ direction is real for each diffraction order.
We can confirm that all diffraction modes in each medium propagate in the same direction,
\begin{align}
  \tan \theta_m^\tau
  \equiv 
    \frac{K_{\mathbf{k}_m}^\tau}{k_{x,m}}
  = \frac{\sqrt{\epsilon^\tau}}{c}
  \operatorname{Re}\sqrt{{v_\mathrm{ph}}^2 - \left(\frac{c}{\sqrt{\epsilon^\tau}}\right)^2},
\end{align}
Note that the far right hand side does not depend on the diffraction order $m$.
Therefore,
different diffraction orders with different frequencies $\omega_m = m\Omega$ are superposed to form pulses.
This feature is more significant in the dielectric side (see the cross section plots in \figref{fig:field}).

What we have observed here is radiation from the virtually moving dipoles induced by the electrostatic input on the interface.
The evenly spaced, moving dipoles collectively emit radiation in the far field,
and thus the field pattern is a plane wave.
We can also mathematically understand the radiation as emission from the source term on the right hand side of Eq. \eqref{eq:bc},
which is induced by the temporal modulation of the interface ($\dot{a}_\mathbf{x} \neq 0$).
Since the strength of the source is proportional to the temporal modulation frequency,
the emitted field is stronger as the modulation speed $v_\mathrm{ph}$ increases in \figref{fig:field}.

\textit{Conclusions.---}%
In this Letter,
we have proposed a mechanism for the \v{C}erenkov radiation in a system which consists of a single interface.
The spatiotemporal modulation of its interface realises the interface profile of a travelling wave.
By applying an electrostatic field,
electric dipoles are induced on the interface,
and move virtually on the interface due to the travelling wave profile.
The profile slides at its phase velocity $v_\mathrm{ph} = \Omega/g$,
where $\Omega$ and $g$ are independent modulation parameters, 
temporal and spatial modulation frequencies.
Thus, the sliding speed of the profile $v_\mathrm{ph}$ is not limited by the speed of light.
If the profile velocity and hence that of the induced dipoles is faster than light,
they emit \v{C}R.

As for the experimental implementation,
the key is the spatiotemporal modulation of the interface between two media.
One can use acoustic techniques to spatiotemporally modulate the surfaces of materials.
According to the studies in acoustic communities \cite{chenu1994giant, mante2010generation},
surface displacement of the order of $10\ \mathrm{nm}$ and ultrafast modulation up to $1\ \mathrm{THz}$ are achievable through acoustic-optical modulation.
Using soft materials such as liquids is one way to generate large surface displacement up to micrometer scale 
\cite{issenmann2006bistability, issenmann2008deformation, rambach2016visualization}.

Another possibility is electrostatic modulation of permittivities of atomically thin materials placed at interfaces.
Those thin materials can be regarded as infinitely thin sheets with finite conductivities \cite{gonccalves2016introduction}.
Recently, several studies have revealed that the permittivity profile of graphene can be electrically modulated 
\cite{galiffi2019broadband, galiffi2020wood, kashef2020multifunctional}.
The modulation can be performed at high frequency in space and time so that graphene is another candidate to realise our proposed effects
although we should take it into consideration that the optical response is dominated by two dimensional Dirac electrons \cite{gonccalves2016introduction}.

\begin{acknowledgments}
  D.O. is funded by the President's PhD Scholarships at Imperial College London.
  K.D. and J.B.P. acknowledges support from the Gordon and Betty Moore Foundation.
\end{acknowledgments}

\end{document}